# Anomalous transport in fractal media with randomly inhomogeneous diffusion barrier

Olga A. Dvoretskaya and Peter S. Kondratenko

*Nuclear Safety Institute, Russian Academy of Sciences,*
*52 Bolshaya Tul'skaya St., 115191 Moscow, Russia*

*Moscow Institute of Physics and Technology (State University),*

*9 Institutskii per., Dolgoprudny, RU-141700 Moscow Region, Russia*
Phone: +7(495) 955 2291, E-mail: kondrat@ibrae.ac.ru

We investigate a contaminant transport in fractal media with randomly inhomogeneous diffusion barrier. The diffusion barrier is a low permeable matrix with extremely rare high permeability pathways (punctures). At times, less than a characteristic matrix diffusion time, the problem is effectively barrier-free with an effective source acting during the time $t_{eff} \ll t$. The punctures result in a precursor contaminant concentration at short times and additional stage of the asymptotic concentration distribution at long times. If the size of the source surface area is large enough, the barrier can be considered as statistical homogeneous medium; otherwise, strong fluctuations occur.

## 1. INTRODUCTION

Anomalous transport in highly heterogeneous media has been a subject of intensive research over the last decades [1-3]. The anomalous diffusion is characterized by the power law dependence of the contaminant plume size $R$ on time $t$

$$R(t) \propto t^{\gamma}, \qquad (1)$$

where $\gamma \neq 1/2$.
So, we observe superdiffusion and subdiffusion for $\gamma < 1/2$ and $\gamma > 1/2$, correspondingly. Subdiffusion occurs in the transport of the charge in disordered semiconductors [4-8] proteins and lipids in the cell membrane [9-13], contaminant particles in porous media [14-15]. In turn, superdiffusion behavior is exhibited by bacteria [16-17], atoms and atomic clusters on metal surfaces [18,19], contaminant particles in heterogeneous rocks [20,21].

Most models of anomalous transport models consider a contaminant source located either in a highly permeable subsystem or at an averaged position (See [22-23]). Because of the practical application, it is worthwhile to consider the problem, where the source is surrounded by the low-permeable diffusion barrier. The influence of the homogeneous diffusion barrier on contaminant transport in regularly and randomly heterogeneous media has been investigated in [24] and [25], correspondingly.

The purpose of the present paper is to analyze contaminant transport in fractal media with the randomly-inhomogeneous diffusion barrier. This barrier consists of the low permeable matrix penetrated by punctures, which have high permeability. We show that the presence of such punctures causes a specific behavior of the contaminant distribution in the main body as well as at the large distances from the source.

The paper is organized as follows. In Sec. II, we formulate the problem and find the relation between the problem of the diffusion barrier and the barrier-free problem. In Sec. III, we analyze the time evolution of the effective source power. In Sec. IV we study the behavior of the main body contaminant concentration. We find asymptotic concentration distribution (at large distances from the source) in Sec. V. Finally, we summarize the results in Sec. VI.

## 2. PROBLEM FORMULATION AND BASIC RELATIONS

The model of the fractal medium with a diffusion barrier is shown in Fig.1. A contaminant source (**S**) is surrounded by a near-field (**N**) and which, in turn, is surrounded by a far-field (**F**), filling the rest of the space. Boundaries between **F** and **N** and between **N** and **S** are concentric spheres. The **N**-**F** boundary radius $a$ is large compared to the radius of the source $a_s$

$$a \gg a_s \tag{2}$$

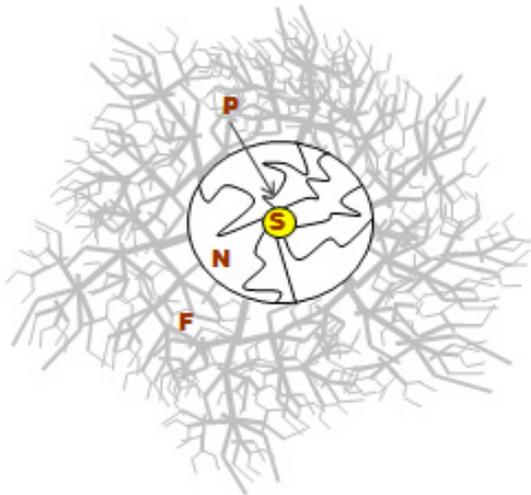

FIG. 1. Schematic sketch of the problem. **S** – source, **N** – near-field, **P** - punctures, **F** – far-field.

The near-field contains a low permeable matrix penetrated by punctures. The punctures are isolated pathways with high permeability, which are randomly distributed over the area. They have tortuous structure and connect the source with the **NF** boundary. The diffusion coefficient in the puncture $D$ is much greater than the diffusion coefficient in the matrix $d$

$$D \gg d \tag{3}$$

Contaminant transport in the far-field (**F**) is caused by advection throughout a disordered fractal system of fractures. The outer radius of the near-field $a$ is of the order of the lower limit of the fractal scaling. So the low permeability **N**-field acts as a diffusion barrier.

The contaminant particles located inside the fracture systems (**F**) are referred to as active. Our goal is to analyze the distribution of active particles at $r \gg a$. At large distances the near-field can be treated as a point source of contaminants with the time-dependent effective source power $Q(t)$ defined by the contaminant flux at **NF** boundary. The contaminant transport mechanism in the far-field is significantly faster than in the near-field. So, we use the following condition for the contaminant concentration at **NF** boundary:

$$c(r,t)\big|_{r=a} = 0 \qquad (4)$$

for the contaminant concentration at **NF** boundary.

Since the punctures are extremely rare, it is reasonable to suppose that a total number of contaminant particles reaching the **NF** boundary is governed mostly by matrix diffusion. However at short times, the contribution of punctures to the effective source power is dominant due to fast diffusion throughout the punctures (see Eq. (3)). So we observe a precursor of the concentration. In accordance with the above, the effective source power can be given as the sum

$$Q(t) = Q_m(t) + Q_p(t), \qquad (5)$$

where subscripts $m$, $p$ denote the matrix and puncture contribution correspondingly.

Also, the concentration in the near- and far- field can be written as

$$c = c_m + c_p \qquad (6)$$

At $t = 0$, the contaminant particles are localized in the source **S**, and their initial total number is denoted by $N_0$.

The initial concentration of contaminants outside of source is zero. If diffusion inside the source is much faster than diffusion inside of matrix, the contaminant concentration is almost homogeneous inside the source, and the concentration at the boundary ($r = a_s$) satisfies the conditions

$$c_m(a_s,t) = \frac{N_0}{V_s} + \frac{d}{3a_s}\int_0^t dt' \frac{\partial c_m(r,t')}{\partial r}\bigg|_{r=a_s+0}, \quad V_s = \frac{4\pi}{3}a_s^3; \quad (7)$$

$$c_m(a_s,t) = c_p(a_s,t).$$

Here, we neglect the flux through the punctures at the boundary $r = a_s$.

In order to describe the contaminant transport in the far-field, we use the isotropic random advection model [26] described by the transport equation

$$\frac{\partial c_i}{\partial t} + div(\vec{v}c_i) = Q_i(t)\delta(\vec{r}); \; i = m, p \quad (8)$$

where advection velocity $\vec{v}(\vec{r})$ is a random function of coordinates obeying the conditions: $div(\vec{v}) = 0$ (incompressible flow) and $<\vec{v}(\vec{r})> = 0$. The sign $<...>$ represents the averaging over the entire ensemble of realizations. Since the fracture system possesses fractal properties, velocity correlations are followed by power law decay (long-range correlations) at large distances, and the two-point velocity correlation function has a form

$$K_{ij}^{(2)}(\vec{r}) \equiv <v_i(\vec{r}_1)v_j(\vec{r}_2)> \sim V^2 (a/|\vec{r}|)^{2h}, \quad |\vec{r}| \gg a \quad (9)$$

with $\vec{r} = \vec{r}_1 - \vec{r}_2$. The term $V^2$ is a characteristic value of $K_{ij}(\vec{r})$ at $|\vec{r}| \sim a$, $h$ is a scaling dimension of velocity fluctuations $v(\vec{r})$. Therefore, $K_{ij}^{(2)}(\vec{r})$ is the scale-invariant function that is

$$K_{ij}^{(2)}(\lambda\vec{r}) = \lambda^{-2h}K_{ij}^{(2)}(\vec{r}) \quad (10)$$

where $\lambda$ is a arbitrary dimensionless positive number. The higher-order correlation functions have similar property.

Taking into account linearity of the problem, the ensemble averaged contaminant concentration $<c_i(\vec{r},t)>$ may be represented as

$$<c_i(\vec{r},t)> = \int_0^t dt' Q_i(t-t') G(\vec{r},t'), \quad (11)$$

where $G(\vec{r},t)$ is the Green's function of the barrier-free problem.

We analyze the transport of active particles for both concentration components ($i = m, p$) in terms of the total number of active particles

$$N_i(t) = \int d\vec{r} < c_i(\vec{r}, t) >, \tag{12}$$

contaminant plume size $R_i(t)$ and $R(t)$ given by

$$R_i^2(t) = \frac{1}{N_i(t)} \int d\vec{r} \cdot < c_i(\vec{r}, t) > r^2, \tag{13}$$

$$R^2(t) = \frac{R_m^2 N_m + R_p^2 N_p}{N(t)}, \tag{14}$$

where
$$N(t) = N_m(t) + N_p(t) \tag{15}$$

and asymptotic concentration distribution at large distances

$$< c_i(\vec{r}, t) > \propto \exp\{-\Phi_i(\vec{r},t)\}, \quad r \gg R_i(t). \tag{16}$$

Note that the isotropic random advection model considered in Ref. [26], corresponds to the barrier-free problem (i.e. without diffusion barrier) and the following results were obtained.

The contaminant plume size $R_*(t)$ is given by

$$R_*(t) \sim \left(a^{\frac{1-\gamma}{\gamma}} Vt\right)^\gamma, \tag{17}$$

where

$$\gamma = \begin{cases} (1+h)^{-1}, & h < 1 \\ 1/2, & h > 1 \end{cases} \tag{18}$$

Thus, for $h > 1$ the superdiffusion is observed, whereas for $h < 1$ - classical diffusion.
The Green's function $G(\vec{r},t)$ behaves as [see [26]]:

$$G(r,t) = (R_*(t))^{-3} F(\varsigma), \quad \varsigma = \frac{r}{R_*(t)};$$
$$F(\varsigma)\big|_{\varsigma \sim 1} \sim 1, \quad F(\varsigma)\big|_{\varsigma \gg 1} \ll 1. \tag{19}$$

For $r \gg R_*(t)$, the above expression takes the form

$$G(r,t) \propto \exp\{-\Phi_*(r,t)\},$$

$$\Phi_*(r,t) = \left(\frac{r}{R_*(t)}\right)^{\frac{1}{1-\gamma}}, \tag{20}$$

where $\gamma$ is given by Eq. (18). Note that the subscript $_*$ in Eqs. (17)-(20) denotes the quantities obtained in the barrier-free problem.

## 3. EFFECTIVE SOURCE POWER

Contaminant transport in the matrix is described by the diffusion equation

$$\left\{\frac{\partial}{\partial t} - d\Delta\right\} c_m(\vec{r},t) = 0 \tag{21}$$

with boundary conditions (4), (7).

Solving this equation and substituting $c_m(r,t)$ into the first Fick's law

$$Q_m(t) = -4\pi a^2 d \left.\frac{\partial c_m(r,t)}{\partial r}\right|_{r=a},$$

we obtain

$$Q_m(t) = \frac{N_0}{4t_m} H\left(\frac{t}{4t_m}\right), \quad t_m = \frac{a^2}{4d};$$

$$H(w) = \int_{y_0-i\infty}^{y_0+i\infty} \frac{dy}{2\pi i} \frac{\sqrt{y} e^{wy}}{sh(\sqrt{y})}, \quad \text{Im}\, y_0 = 0,\, y_0 > 0 \tag{22}$$

Thus, for short and long times we get [26]

$$Q_m(t) \cong 4N_0 \sqrt{\frac{t_m^3}{\pi t^5}} \exp\left(-\frac{t_m}{t}\right), \quad t \ll t_m;$$

$$Q_m(t) \cong N_0 \frac{\pi^2}{2t_m} \exp\left(-\frac{\pi^2}{4}\frac{t}{t_m}\right), \quad t \gg t_m. \tag{23}$$

<u>Puncture contribution</u>, $Q_p(t)$

First, we calculate a contribution of the individual puncture to the effective source power, i.e. diffusive flux of the contaminants arriving at the **NF** boundary ($r = a$) $q(l,t)$ by diffusion throughout the individual puncture. The puncture is a quasi- one-dimensional object of length $l$ and small cross-sectional area $s_0$

$$s_0 \ll l^2. \tag{24}$$

So the concentration distribution inside the puncture satisfies the equation

$$\left(\frac{\partial}{\partial t} - D\frac{\partial^2}{\partial x^2}\right) c_p(x,t) = 0 \qquad (25)$$

with boundary conditions $c_p(0,t) = c_m(a_s,t)$, $c_p(l,t) = 0$. Variable $x$ is a coordinate along the puncture.

Substituting the solution of above equation into the expression $q = -s_0\left[(\partial/\partial x)c_p(x,t)\right]_{x=l}$, we obtain

$$q(l,t) = \frac{6\sqrt{\pi} N_0 s_0}{S^{3/2}} \frac{D}{l} \int_{p_0-i\infty}^{p_0+i\infty} \frac{dp}{2\pi i} e^{pt} \frac{1}{p\left[1 + \frac{3}{4pt_s}\left(1 + 2\sqrt{pt_s}\right)\right]} \frac{\left(2\sqrt{pt_l}\right)}{sh\left(2\sqrt{pt_l}\right)}, \qquad (26)$$

Here $p_0$ is a real positive number, i.e. $\operatorname{Im} p_0 = 0$, $p_0 > 0$. $S = 4\pi a_s^2$ is the surface area of the source, and the characteristic times $t_l$ and $t_s$ are defined as

$$t_l = \frac{l^2}{4D}, \quad t_s = \frac{S}{16\pi d} = \left(\frac{a_s}{a}\right)^2 t_m. \qquad (27)$$

Depending on the relation between $t_l$ and $t_s$ and $t$ the contribution of the single puncture takes the form

$$q(l,t) \cong \frac{12\sqrt{\pi} N_0 s_0}{S^{3/2}} \sqrt{\frac{D}{\pi t}} B\left(\frac{t}{\sqrt{t_l t_s}}\right) \exp(-t_l/t), \quad t \ll t_l;$$

$$B(x) = \left[1 + \frac{3}{2}x\left(1 + \frac{x}{2}\right)\right]^{-1}. \qquad (28)$$

$$q(l,t) \cong \frac{6\sqrt{\pi} N_0 s_0}{S^{3/2}} \frac{D}{l}, \quad t_l \ll t \ll t_s. \qquad (29)$$

$$q(l,t) \cong \frac{N_0 s_0}{\left(4\pi d t\right)^{3/2}} \frac{D}{l}, \quad t \gg t_s. \qquad (30)$$

Thus, the flux of contaminants from the puncture depends exponentially on time in the beginning, i.e. at $t \ll t_l$. At $t_l \ll t \ll t_s$, the flux saturates and, decays as a power-law decay at $t \gg t_s$. The sketch of the function of $q(l,t)$ is shown in Fig. 2.

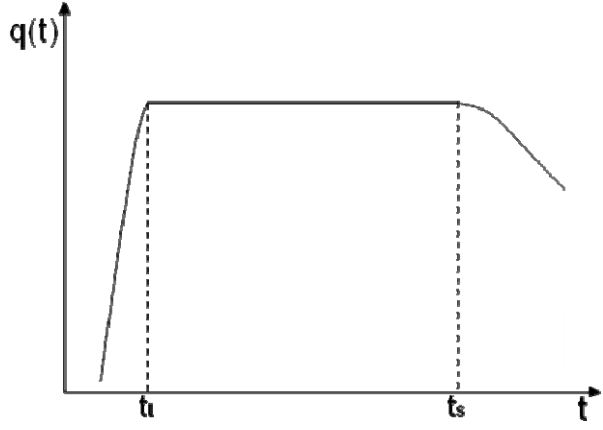

FIG 2. Time evolution of the individual puncture contribution to the effective source power (schematically)

To calculate the total contribution of all possible punctures to the effective source power, which we referred to as the total puncture contribution henceforth. We use the approach developed by Raich and Ruzin [27] to investigate the conductance of the tunnel junction in amorphous semiconductors. We assume that the length of the puncture $l$ varies randomly, i.e. $l = au$, where $u$ is a dimensionless random variable. Other puncture characteristics such as the cross-section area $s_0$ and diffusion coefficient $D$ are constant. The total puncture contribution can be written as

$$Q_p(t) = S \cdot \int_1^{u_p} du\, \rho(u) q(au,t), \qquad (31)$$

where $\rho(u)$ is the density distribution function of the punctures.

$$\rho(u) = s_0^{-1} \exp\left[-\Omega(u)\right]. \qquad (32)$$

The argument of the exponent in Eq. (32) satisfies the following conditions (see [27])

$$\Omega(u) \gg 1, \quad \partial\Omega/\partial u < 0, \quad \partial^2\Omega/\partial u^2 > 0. \qquad (33)$$

The lower limit of integration in Eq. (31) $u = 1$ is determined by the shortest path between the source and the boundary. The upper limit $u_p = l_p/a$ is determined by the length $l_p$ at which contaminant particles have enough time to leave the punctures due to matrix diffusion (верхний предел определяется из условия, что за время диффузии по проколу примесь может успеть покинуть его, продиффунидировав в матрицу). So we get

$$u_p \sim \frac{\sqrt{s_0}}{a}\sqrt{\frac{D}{d}}. \qquad (34)$$

In accordance with Eqs (28), (29), (31) the behavior of the effective source power is different for three time intervals: 1) $t \ll t_p$, 2) $t_p \ll t \ll t_s$, 3) $t \gg t_s$, where $t_p = \dfrac{a^2 u_p^2}{4D} = \dfrac{s_0}{4d}$. Note that $t_p \ll t_s \ll t_m$. Further we consider the behavior of $Q_p(t)$ in each of these intervals.

### $t \ll t_p$

The integrand in (31) is the product of two rapidly changing functions: one of them $\rho(u)$ increases with $u$, while the other one $q(au,t)$ decreases. So, we perform the integration by means of the saddle point method. Using Eqs. (28), (31), we obtain

$$Q_p(t) \simeq \frac{12 N_0 \sqrt{\pi}}{\sqrt{S}} \sqrt{\frac{D}{t\Omega''(u_{opt}) + 2t_a}} \exp\left(-\Omega(u_{opt}) - \frac{t_a}{t} u_{opt}^2\right) \tag{35}$$

with $t_a = \dfrac{a^2}{4D}$.

The optimum value of $u$ corresponding to the optimum punctures $u_{opt}$ is found from the saddle point equation

$$\left.\frac{\partial \Omega(u)}{\partial u}\right|_{u=u_{opt}} + \frac{2 t_a u_{opt}}{t} = 0 \tag{36}$$

The relation (35) is valid where the number of optimum punctures located at the surface area of the source is sufficiently large

$$\frac{S}{s_0} \exp\left[-\Omega(u_{opt})\right] \gg 1 \tag{37}$$

Otherwise,

$$\frac{S}{s_0} \exp\left[-\Omega(u_{opt})\right] \ll 1 \tag{38}$$

and the effective source power is determined by the "typical punctures" (See [27]), the number of which is of the order of unity. The value of $u$ for typical punctures ($u_f$) is found from the relation

$$\frac{S}{s_0} \exp\left[-\Omega(u_f)\right] = 1 \tag{39}$$

In this case the upper limit of the integration (31) should be replaced by $u_f$. Since the integral (31) converges near $u_f$, we get

$$Q_p(t) = \frac{12\sqrt{\pi} N_0 s_0}{S^{3/2}} \sqrt{\frac{D}{\pi t}} \frac{1}{\Omega'(u_f) + 2 u_f t_a / t} \exp\left(-\frac{t_a}{t} u_f^2\right), \tag{40}$$

This expression is valid for $u_f < u_p$. If $u_f > u_p$, the integral converges near the upper limit ($u_p$). So $u_f$ should be replaced by $u_p$ in the above relation.

Thus, expression (35) is valid for the large surface area of the source $S > S_{cr}$, and (40) - for the small ones $S < S_{cr}$. The critical value of the surface area of the source $S_{cr}$ is

$$S_{cr} = s_0 \exp\left[\Omega(u_{opt})\right] \tag{41}$$

Taking into account (35), (40) the effective source power can be expressed as

$$Q_p(t) \propto \exp(-F(t)), \tag{42}$$

with

$$F(t) = \begin{cases} \Omega(u_{opt}) + \dfrac{t_a}{t} u_{opt}^2, & S > S_{cr} \\ \dfrac{t_a}{t} u_k^2, & S < S_{cr} \end{cases}, \tag{43}$$

where subscript $k = p, f$ for $u_f > u_p$ and $u_f < u_p$, respectively.

$\underline{t_p \lll t_s}$

Since the integrand (24) is a fast growing function, the integral converges near the upper limit. So the effective source power takes the form

$$Q_p(t) = Q_p^{(st)},$$
$$Q_p^{(st)} \simeq \frac{3N_0}{\sqrt{t_s t_p}} \frac{D}{d} \frac{1}{|\Omega'(u_p)|} \exp\left[-\Omega(u_p)\right]. \tag{44}$$

Thus, $Q_p$ does not depend on time.

$\underline{t \ggg t_s}$

Here, the effective source power has power-law decay on time

$$Q_p(t) = \frac{Q_p^{(st)}}{3} \left(\frac{t_s}{t}\right)^{3/2}. \tag{45}$$

Now, let us calculate the total number of contaminant particles delivered to the **NF** boundary by punctures. Using Eqs. (12) and (31), we obtain

$$N_p(\infty) \equiv \int_0^\infty dt Q_p(t) = N_0 \frac{D}{2\sqrt{\pi}d} \frac{\sqrt{S}}{a} \int_1^{u_p} \frac{du}{u} \exp[-\Omega(u)] \simeq$$
$$\simeq N_0 \sqrt{\frac{Dt_s}{dt_p}} \frac{1}{|\Omega'(u_p)|} \exp[-\Omega(u_p)] \tag{46}$$

The applicability condition of the model under consideration requires $N_p(\infty)$ being small compared to $N_0$.

$$\sqrt{\frac{Dt_s}{dt_p}} \frac{1}{|\Omega'(u_p)|} \exp[-\Omega(u_p)] \ll 1. \tag{47}$$

It is interesting to estimate the relative statistical spread of the effective source power. The spread is caused by strong spatial fluctuations of the barrier characteristics and defined by

$$\Delta = \frac{\sqrt{<(\delta Q_p)^2>}}{Q_p}, \tag{48}$$

where $<...>$ denotes the ensemble averaging and $<(\delta Q_p)^2> = <Q_p^2> - <Q_p>^2$. Similar calculations [27] found the relative statistical spread of the conductivity:

$$\begin{array}{ll} \Delta \ll 1, & S > S_{cr}, \\ \Delta \sim 1, & S < S_{cr}, \\ \Delta \gg 1, & S \ll S_{cr}. \end{array} \tag{49}$$

Here, $S_{cr}$ is given by Eq. (41) for $t \ll t_p$, and $S_{cr} = s_0 \exp[\Omega(u_p)]$ for $t \gg t_p$.

It follows from Eq. (11) that the statistical spread of the concentration is given by Eq. (49) as well. Thus, the relative statistical spread of the effective power and concentration is negligible, where $S > S_{cr}$, and the medium is homogeneous in average (statistical homogeneity case). In contrast, the statistical spread is large, where $S < S_{cr}$, and strong fluctuations are observed At intermediate values of $S$, it is of order of unity.

Notice that for $t \gg t_p$ the effective power does not depend on whether the source surface area is more or less than its critical value, but the relative statistical spread still depends on.

### 4. TRANSPORT REGIMES

Transport regime in far field is determined by the total number of active particles $N_i(t)$ given by Eq. (12) and the contaminant plume size $R_i(t)$ defined by Eq. (13). We substitute Eq. (11) in Eqs. (12),(13) and find

$$N_i(t) = \int_0^t dt' Q_i(t'), \tag{50}$$

$$R_i(t) = \frac{1}{N_i} \int_0^t dt' R_*(t') Q_i(t-t'). \tag{51}$$

where the identity $\int d^3 r \, G(r,t) = 1$ was used.

First, we consider transport characteristics related to the matrix ($i = m$). At short times $t \ll t_m$ the dominant contribution to the integrals in Eqs (46), (47) comes from the vicinity of the lower integration limit $t' \ll t_m$. Therefore, we use the expression for $Q_m(t-t')$ obtained from Eq. (17) by expanding exponent to the first order:

$$Q_m(t-t') \cong Q_m(t) \exp\left(-\frac{t'}{t_{eff}^{(m)}(t)}\right), \quad t_{eff}^{(m)}(t) = \frac{t^2}{t_m} \tag{52}$$

Substituting this equation into Eqs. (50), (51), and using Eq. (17), we get

$$\begin{aligned} N_m(t) &\cong N_0 \cdot 4 \sqrt{\frac{t_m}{\pi t}} \exp\left(-\frac{t_m}{t}\right), \\ R_m(t) &\simeq R_*\left(t_{eff}^{(m)}\right) \sim \left(a^{\frac{1-\gamma}{\gamma}} V \frac{t^2}{t_m}\right)^{\gamma} \end{aligned} \qquad t \ll t_m \tag{53}$$

For $t \gg t_m$ integrals in Eq. (11) rapid converge at $t - t' \sim t_m$. So using Eqs. (50), (51), we find

$$\begin{aligned} N_m &\simeq N_0, \\ R_m(t) &\sim R_*(t) \sim \left(a^{\frac{1-\gamma}{\gamma}} V t\right)^{\gamma}; \end{aligned} \qquad t \gg t_m. \tag{54}$$

Note that Eqs. (53), (54) were found in [26].

Similarly, we find quantities $N_p(t)$ and $R_p(t)$, which are the transport characteristics related to punctures contribution.

$\underline{t \ll t_p}$

$$\begin{aligned} N_p(t) &\simeq Q_p(t) \cdot t_{eff}(t), \\ R_p(t) &\sim R_*\left(t_{eff}\right). \end{aligned} \tag{55}$$

Here, $Q_p(t)$ is given by Eqs. (35) and (40) for $S > S_{cr}$ and $S < S_{cr}$, respectively. The effective time is

$$t_{eff}(t) = \left|\frac{dF(t)}{dt}\right|^{-1}, \qquad (56)$$

where function $F(t)$ is defined by Eq. (43).

$\underline{t_p \ll t \ll t_s}$

$$N_p(t) \simeq tQ_p^{(st)},$$
$$R_p(t) \sim R_*(t). \qquad (57)$$

with $Q_p^{(st)}$ given by Eq. (44).

$\underline{t \gg t_s}$

$$N_p(t) \simeq N_p(\infty),$$
$$R_p(t) \sim R_*(t). \qquad (58)$$

Comparing Eqs. (55)-(58) with Eq. (53), we conclude that for $t \ll \dfrac{t_m}{\Omega(u_p)}$ the punctures contribution to the total number of active particles is dominating and $N_p(t) \gg N_m(t)$. The evolution of $N(t)$, given by (15), is schematically depicted in Fig. 3.

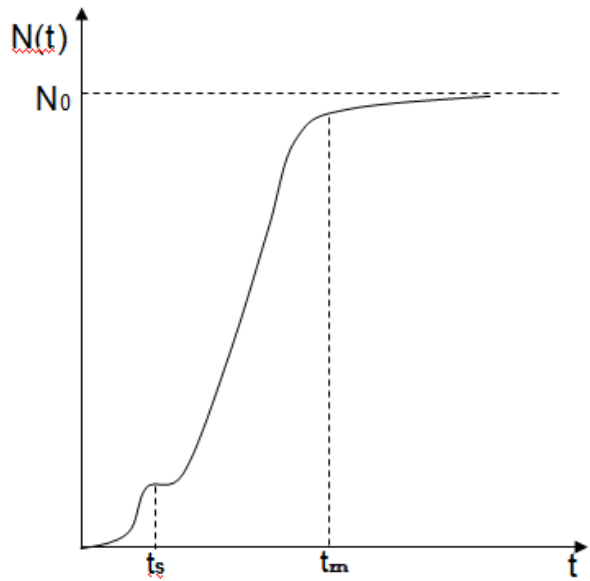

FIG.3. Evolution of the active particles number $N(t)$ (schematically).

The evolution of the contaminant plume size $R(t)$ given by Eq. (14) is schematically depicted in Fig. 4.

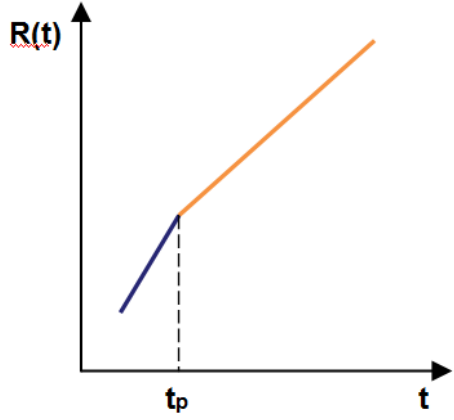

FIG.4. The evolution of the contaminant plume size $R(t)$ (schematically).

## 5. ASYMPTOTIC CONCENTRATION DISTRIBUTIONS

<u>Matrix contribution</u>

Substituting Eqs. (19) and (23) in Eq. (11), we obtain

$$<c_m(\vec{r},t)> \propto \int_0^t dt' \exp\left(-\frac{t_m}{t-t'}\right)\exp\left(-\Phi_*(r,t')\right). \tag{59}$$

At short times $t \ll t_m$, the first exponent in the integrand rapidly decreases with $t'$, whereas the second one increases. Therefore, the integrand has a sharp maximum, and the saddle-point method is applied. The result of the integration depends on the ratio between $\frac{t_m}{t}$ and $\Phi_*(r,t)$. If $\frac{t_m}{t} \gg \Phi_*(r,t)$ the saddle point $t'_0$ is much less than $t$. The saddle-point equation is

$$\frac{t_m}{t^2} + \dot{\Phi}_*(r,t'_0) = 0. \tag{60}$$

Here and below, we denote the derivative with respect to time by the dot.

Taking into account Eq. (20) and solving Eq. (60), we find the exponent in the asymptotic concentration distribution from Eq. (16)

$$\Phi_m(r,t) = \frac{t_m}{t} + \frac{1}{\gamma^\gamma(1-\gamma)^{1-\gamma}}\frac{r}{R_*(t_{eff}(t))}, \quad \Phi_*(r,t)\frac{t}{t_m} \ll 1. \tag{61}$$

If $\Phi_*(r,t) \gg \frac{t_m}{t}$, then $t'_0$ is close to $t$. So, the saddle-point equation takes the form

$$\left|\dot{\Phi}_*(r,t)\right| + \frac{t_m}{(t-t'_0)^2} = 0. \tag{62}$$

Thus, we have

$$\Phi_m(r,t) = 2\sqrt{\frac{\gamma}{1-\gamma}\frac{t_m}{t}\Phi_*(r,t)} + \Phi_*(r,t), \quad \Phi_*(r,t)\frac{t}{t_m} \gg 1. \tag{63}$$

This expression is also valid for $t \geq t_m$.

Note that for $t \ll t_m$, the asymptotic concentration (concentration tails) has two stages, whereas for $t \gg t_m$ only one stage. The first terms in Eqs. (61), (63) determine the number of active particles forming the asymptotic concentration distribution. In all cases, their number is exponentially small.

### Punctures contribution

Substituting Eqs. (20) and (35) into Eq. (11), we find the asymptotic concentration distribution due to the puncture contribution

$$<c_p(\vec{r},t)> \propto \int_0^t dt' \exp\left(-F(t-t') - \Phi_*(r,t')\right) \tag{64}$$

It is clear that the integration result depends on which exponential factor is faster. So, we consider two distinct cases 1) $\dot{\Phi}_*(r,t) \ll \dot{F}(t)$ and 2) $\dot{\Phi}_*(r,t) \gg \dot{F}(t)$.

#### 1. $\underline{\dot{\Phi}_*(r,t) \ll \dot{F}(t)}$

Using the Taylor expansion foe $F(t-t')$ at $t'=0$, the exponent in Eq. (64) takes the form: $F(t) + |\dot{F}(t)|t' + \Phi_*(r,t')$. We use the saddle-point method to integrate Eq. (64). The saddle-point value $t'_0$ is found from the equation:

$$\dot{\Phi}_*(r,t'_0) = -|\dot{F}(t)|. \tag{65}$$

Thus, the exponent $\Phi_p$ of the asymptotic concentration distribution in Eq. (16) is

$$\Phi_p(r,t) = F(t) + \frac{1}{\gamma^\gamma (1-\gamma)^{1-\gamma}} \frac{r}{R_*(t_{eff}(t))}, \tag{66}$$

#### 2. $\underline{\dot{\Phi}_*(r,t) \gg \dot{F}(t)}$

Substituting $t-t'$ for $t'$ in Eq. (64) and applying the Taylor series expansion for $\Phi_*(r, t-t')$ around the point $t'=0$, we find the exponent in Eq. (20) takes the form

$$\Phi_*(r,t) + |\dot{\Phi}_*(r,t)|t' + F(t').$$

We obtain the saddle-point value $t'_0$ is found from the equation

$$|\dot{\Phi}_*(r,t)| = -\dot{F}(t'_0) \tag{67}$$

Let us assume that the function $F(t)$ has the form $F(t) = \left(\dfrac{t_*}{t}\right)^\beta$. This form is relevant because it obeys conditions (33) for $\Omega(t)$, and the probability of the puncture is extremely low. So the exponent of the asymptotic concentration expansion (16) takes the form

$$\Phi_p(r,t) = \Phi_*(r,t) + (1+\beta)\left[\frac{t_*|\dot{\Phi}_*(r,t)|}{\beta}\right]^{\frac{\beta}{1+\beta}} \tag{68}$$

The dependence of the concentration ($c = c_p + c_m$) dependence on the distance at $r \gg R(t)$ (concentration tail) is schematically shown in Fig. 5. The first two stages of the tail are determined by the concentration behavior of the contaminant particles which arrive come to the far-field from the matrix, and the last stage is formed by the particles delivered by punctures at the earliest times. At long distances, as well as at short times, the effect of punctures is significant for the concentration distribution.

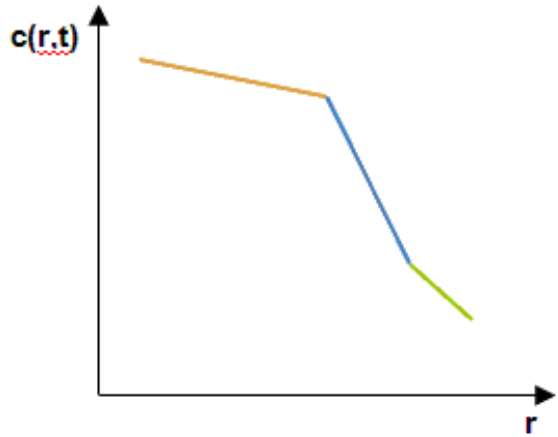

FIG. 5. The asymptotic concentration with distance (schematically).

**CONCLUSION**

We analyzed the contaminant transport in a fractal medium with randomly inhomogeneous diffusion barrier. The barrier consists of the two subsystems: the low permeability matrix and punctures (randomly distributed extremely rare isolated pathways with high permeability which penetrate the matrix).

For the concentration in the main body, we have obtained the following results. At times, less than the characteristic matrix diffusion time $t \ll t_m$, the problem with diffusion barrier is effectively "barrier-free" with an effective source acting during the time $t_{eff} \ll t$. So, the diffusion barrier results in the retardation of the growth of the contaminant plume. The punctures lead to the precursor concentration. Although the number of contaminant particles delivered through the punctures is much smaller than through the matrix, the contribution of the punctures is dominant at short times.

The diffusion barrier causes the modification of the contaminant concentration at large distances (concentration tail). The concentration behavior is governed by the "earliest" particles arriving from the punctures. So, the additional stage (the most remote one) of the concentration tail is observed.

The size of the source surface area greatly affects the contribution of punctures. If the source surface area is large enough, the medium is homogeneous in average (statistical homogeneity case). Hence the relative statistical spread of the effective source power and contaminant concentration is small compared to unity. Otherwise, the strong fluctuations are observed, and the statistical spread is large.

**ACKNOWLEDGMENT**

This work supported by the Russian Foundation of Basic Research (RFBR) under project 08-08-01009a.